# Biocompatible carbon nitride-based light-driven microswimmer propulsion in biological and ionic media with responsive on-demand drug delivery

Varun Sridhar[1], Filip Podjaski[2*], Yunus Alapan[1], Julia Kröger[2,3], Lars Grunenberg[2,3], Vimal Kishore[1], Bettina V. Lotsch[2,3,4*], and Metin Sitti[1,5,6*]

[1] Physical Intelligence Department, Max Planck Institute for Intelligent Systems, 70569 Stuttgart, Germany

[2] Nanochemistry Department, Max Planck Institute for Solid State Research, 70569 Stuttgart, Germany

[3] Department of Chemistry, Ludwig Maximilian University of Munich, Munich, Germany

[4] Cluster of Excellence E-conversion, Lichtenbergstrasse 4, 85748 Garching, Germany

[5] Institute for Biomedical Engineering, ETH Zurich, 8092 Zurich, Switzerland

[6] School of Medicine and College of Engineering, Koç University, 34450 Istanbul, Turkey

* Correspondence to: sitti@is.mpg.de, b.lotsch@fkf.mpg.de, f.podjaski@fkf.mpg.de

**Abstract**

We propose two-dimensional organic poly(heptazine imide) (PHI) carbon nitride microparticles as light-driven microswimmers in various ionic and biological media. Their demonstrated high-speed (15-23 µm/s) swimming in multi-component ionic solutions with concentrations up to 1 M and without dedicated fuels is unprecedented, overcoming one of the bottlenecks of previous light-driven microswimmers. Such high ion tolerance is attributed to a favorable interplay between the particle's textural and structural nanoporosity and optoionic properties, facilitating ionic interactions in solutions with high salinity. Biocompatibility of the microswimmers is validated by cell viability tests with three different cell types and primary cells. The nanopores of the swimmers are loaded with a model cancer drug, doxorubicin (DOX), in high (185%) loading efficiency without passive release. Controlled drug release is reported in different pH conditions and can be triggered on-demand also by illumination. Light-triggered, boosted release of DOX and its active degradation products is demonstrated in oxygen-poor conditions using the intrinsic, environmentally sensitive and light-induced charge storage properties of PHI, which could enable future theranostic applications in oxygen-deprived tumor regions. These organic PHI microswimmers simultaneously solve the current light-driven microswimmer challenges of high ion tolerance, fuel-free high-speed propulsion in biological media, biocompatibility and controlled on-demand cargo release towards their biomedical, environmental and other potential future applications.

**Introduction**

Microswimmers are cell-scale mobile machines that can be self-propelled by converting energy made available to them from their environment or remotely,[1, 2, 3] such as chemical reagents, light, and magnetic or acoustic energy.[4, 5] One of the main targeted applications of microswimmers is local, active, and on-demand delivery of diagnostic or therapeutic cargos, such as drugs, imaging agents, and stem cells, inside the human body and organ-on-a-chip devices.[6, 7, 8] So far, chemical propulsion, such as dissolution of metals (e.g., magnesium in acidic conditions), enzymatic and magnetic propulsion are the most widely used methods in such *in vitro* and *in vivo* biomedical applications.[9] Chemical propulsion is predominantly used for swimming in acidic body fluids, such as inside the stomach and gastrointestinal tract,[9, 10, 11, 12] whereas enzymatic propulsion can be used in other body regions.[13, 14, 15] Both methods need sacrificial and usually toxic agents as fuels to enable swimming, which is disadvantageous for their prolonged use in biological conditions.[16] Light, however, is a viable energy source for the propulsion of microswimmers,[17, 18] both with and without the use of additional fuels, enabling also propulsion control and steering.[19, 20] Moreover, photocatalytic swimmers can exhibit positive and negative phototaxis and gravitaxis,[21] depending on their surface charge,[22] enabling their steering control and targeting to a desired location.[23]

Despite their many advantages over other self-propulsion methods, photocatalytic and chemical propulsion suffer from fundamental propulsion limitations in electrolyte solutions,[18] especially when diffusiophoretic and electrophoretic propulsion modes are operative. This drawback is due to the presence of ions hindering the build-up of concentration gradients of reactants and products involved in self-electrophoresis and self-diffusiophoresis around the swimmer.[24, 25, 26] Although different semi-conducting inorganic photocatalysts[27, 28, 29, 30, 31, 32] have been studied as light-driven microswimmers for potential biological applications[33], donor-free light-driven swimming in ionic media has only been demonstrated with sophisticated geometries in Si and for concentrations below 10 mM. Thus, light-driven swimming in ionic media, such as NaCl, Dubelco's phosphate buffer saline (dPBS), and Dulbeco's modified Engle's medium (DMEM) with concentration levels of over 200 mM present in various biological solutions and cell environments, still remains an unresolved bottleneck. The EI50 number has been introduced previously as a measure of the ionic concentration at which the speed of the microswimmers is reduced by 50%.[34] EI50 is less than 0.1 mM for self-diffusiophoretic and self-electrophoretic swimmers,[35] reaching up to ~4 mM for geometrically optimized systems addressing this problem.[34] For biomedial applications, an EI50 value of more than 100 mM is needed. This limitation is attributed to the presence of a solid surface in the microswimmer, which does not allow fluids or ions to migrate through the swimmer. The presence of salt ions reduces the Debye length of the electrochemical layer formed on the surface of the microswimmers in contact with the solution from hundreds of nm to ~1 nm,[36] thereby collapsing also the ionic gradient around the particle under illumination, which is responsible for the propulsion force. Therefore, commonly, the Debye layer collapse stops the swimming for hard spheres. Besides, most light-driven microswimmers use high concentrations of $H_2O_2$ or alcohols as additional chemical fuels,[37] which are toxic for their potential use in biological applications. The availability of potential biocompatible fuels that are present in large enough quantities has been a pressing bottleneck in implementing light-driven microswimmers in such applications,[18] which require concentrations as high as ~30 mM of biocompatible fuels like glucose for effective light-driven propulsion.[38] Furthermore, taxis-based direction-controlled propulsion as well as controlled cargo uptake and release of active products are highly desired properties of medical microswimmers, which are usually studied and realized separately.[39]

Here, we aim to solve above issues by using highly active (photocatalytically) and porous (texturally and structurally) carbon nitrides ($CN_x$) as light-driven microswimmers. $CN_x$ are a family of organic macromolecular photocatalysts that have gained attention because of their facile synthesis,[40] chemical robustness, absorption of visible light and favorable band positions that enable various photocatalytic redox reactions, such as water splitting.[41] Besides, $CN_x$ are widely used for environmental remediation,[42] sensing and ion pumping.[42,43] [44] Most commonly, a one-dimensional (1D) $CN_x$ called "melon" or "g-$C_3N_4$", is reported. However, in this study we use a recently discovered two-dimensional (2D) carbon nitride species called poly(heptazine imide) (PHI), [45, 46, 47] which hosts hydrated alkali metal ions (typically $K^+$) or protons in its structural nanopores, resulting in an unusual blend of optoelectronic and optoionic properties.[48, 49, 50] PHI does not only show higher hydrogen evolution activity than melon-type $CN_x$,[45, 46, 51] but it is also able to store light-induced electrons for subsequent use through electron-ion interactions.[48, 50, 52] [49] Since PHI microswimmers exhibit not only structural but also textural porosity enabling ion intercalation and permeability that can be driven by light,[43, 47, 48, 49] they are promising platforms for light-driven propulsion[52] and cargo delivery in various biological media. Hence, the development of biocompatible and highly ion-tolerant, non-toxic light-driven microswimmers that can be propelled purely by light in naturally occurring bio-fuels while being able to carry and release a cargo in a controllable fashion is proposed here to solve many fundamental challenges.[49]

## Results

### PHI characterization

The PHI microparticles were obtained by sonication and centrifuge-assisted separation from the original suspension (see Materials and Methods for details). **Figure 1a** shows a scanning electron microscopy (SEM) image of the PHI microparticles with a diameter ranging from 1-5 µm. As they represent agglomerates of smaller primary crystallites, the microswimmers have a slightly irregular spherical shape and textural nanopores occur naturally. These larger voids enable efficient fluid movement and entrapment. **Figures 1b,c** illustrate the schematic and underlying molecular structure of the PHI microswimmers. Structural nanopores in the PHI backbone consisting of imide-bridged heptazine units are filled with hydrated ions that can be exchanged.[47,50] The absorbance spectrum in Figure 1c shows light absorption below 450 nm (band gap of 2.7 eV),[50,48] enabling photocatalysis and, hence, propulsion driven by visible light. The strongly positive valence band position (+2.2 V versus Normal Hydrogen Electrode (NHE)) gives rise to an even stronger oxidative power of the light-generated holes than for melon (**Figure 1d**),[48] providing the thermodynamic driving force for various photocatalytic oxidation reactions including water oxidation, which are often the bottleneck for light-generated charge extraction and photocatalytic propulsion.[52]

### Light-induced swimming in different biological media

In order to demonstrate the mechanism of propulsion of the PHI microswimmers, the light from a photodiode was focused with an intensity of 1.9 W/cm$^2$ at 385 nm on the particle chamber. The mean speed of the illuminated microswimmers in deionized (DI) water was 24.2 ± 1.9 µm/s (**Figure 2a** and **Movie S1**). In this case, their photocatalytically induced swimming is attributed to water oxidation and reduction of dissolved oxygen on the illuminated PHI hemisphere.[52] To explore and determine the influence of different ionic and biological constituents on the propulsion efficiency of the PHI microswimmers, different pH neutral buffers and relevant biological media were tested (see **Supplementary Note 1** for the ingredients and concentrations of all studied liquids). In brief, dPBS was used as the buffer solution for cell washing, which contains NaCl, KCl, $Na_2HPO_4$, and $KH_2PO_4$ at ca. 10 g/L (~150 mM) in total. DMEM, commonly used to culture cells, contains the same salts (~160 mM), amino acids (ca. 2 g/L, ~10 mM), and trace amounts of vitamins and glucose (4.5 g/L, ~25 mM), some of which can be used as reducing agents and, hence, hole extraction fuel to power light-driven microswimmers.[53] Under illumination, the mean speed of the microswimmers in dPBS was 19.0 ± 3.3 µm/s (21% slower than in DI water) and 23.7 ± 2.6 µm/s in DMEM (comparable to DI water), as seen in **Figure 2a**. The ions present in dPBS (~150 mM) hence only have a relatively small influence on the speed of PHI, whereas the additional components in DMEM do not hamper the efficiency of photocatalytic reactions being responsible for the propulsion (**Movie S2**). To emulate a realistic cell environment, the growth-supplementing medium Fetal Bovine Serum (FBS) containing complex proteins and other components[54] were added to DMEM. The illuminated PHI microswimmers still moved with a mean speed of 12.5 ± 3.1 µm/s in DMEM medium with 10% FBS, hence being 47% slower than in DMEM alone. This decrease in speed may be ascribed to the viscosity change and deposition of protein and other components on PHI, thus blocking surface reactions partially.[51] Additionally, a high concentration (1 M) of sodium phosphate buffer, an important component of DMEM, was also used to test the propulsion of PHI microswimmers. We observed a mean speed of 9.4 ± 1.7 µm/s, which is 54% lower than in the 160 mM salt containing DMEM. These findings show that PHI microswimmers can be used efficiently in many realistic biological environments and in salt solutions at concentrations even beyond the biological fluids.

Active motion in heterogeneous and complex biological fluids, such as blood, is also crucial for future biomedical applications in the human body. To test the propulsion capability of the PHI microswimmers in blood, they were mixed with blood cells with a 25% hematocrit (in DMEM medium), which is in the physiological range. When illuminated with UV light (385 nm), the PHI microswimmers also moved in the diluted red blood cell (RBC) solution. Auto-fluorescence of the PHI microswimmers, i.e., the intrinsic emission of PHI without any fluorescent markers, enabled their label-free imaging and detection even in the optically dense RBC solution. The mean speed of the PHI microswimmers in the dilute RBC solution, where viscosity remains the same in comparison with pure DMEM solution,[55] was 20.2 ± 0.8 µm/s, also comparable to their mean speed in DMEM. The RBCs in the solution were also observed to be moving in the same direction, which appears to be caused by the fluidic flow arising from the collective propulsion of PHI microswimmers (**Supplementary Note 2**). Of note, the PHI microswimmers were not able to swim in whole blood, which could be ascribed to the increased viscosity and heterogeneous non-Newtonian nature, making it impossible to swim effectively. A comparison of the available literature and the current work on ionic swimming is shown in **Table S1**. For better comparison, a discussion of the wavelength and intensity dependence of the propulsion can be found in **Supplementary Note 2**, **Figures S1-S2** and **Tables S2-S4**. In DMEM, light-driven propulsion was efficient and proportional to the intensity in the UV spectrum and with blue light at 415 nm, as shown in **Figure 2b**.

**Ionic tolerance for light-induced propulsion with NaCl**

To better understand the ionic tolerance of PHI, the microswimmers were exposed to various NaCl concentrations and compared to the "sister material" melon under 385 nm illumination, as shown in **Figure 2c**. In comparison with DI water (24.2 ± 1.9 µm/s), the PHI swimmers showed a 25% decreased speed (18.1 ± 3.4 µm/s) at 1 mM NaCl, with no obvious further decrease in speed observed until 100 mM (18.6 ± 3.4 µm/s). At 1 M NaCl, they showed a further decrease (12%) in speed (16.1 ± 3.9 µm/s) as seen on **Movie S3**. Hence, the speed was reduced only by 33% with respect to DI water and the characteristic EI50 is not yet reached; thus, PHI's ion tolerance surpasses all reported light-driven microswimmers by two orders of magnitude, while maintaining fast light-induced propulsion above 100 mM salt concentrations.

We attribute the constant speed between 1 mM and 100 mM to a constantly high ionic mobility between the PHI microswimmer and ionic environment, which only becomes limiting at higher concentrations. To further investigate the effect of textural porosity on the high ion tolerance, PHI and melon were tested under identical conditions. While both melon and PHI exhibit textural porosity (see **Figures 1a, S7** and **S8** for SEM and BET measurements), melon has no structural porosity as it consists of close-packed 1D heptazine imide chains, hence enabling only surface ionic interactions. The mean speed of the melon microswimmers under UV illumination was 11.7 ± 2.9 µm/s in DI water (43% lower than PHI). At 10 mM NaCl, the propulsion speed decreased weakly (9%), with EI50 being reached at ~100 mM (6 ± 1.3 µm/s). These findings show that also this 1D form of carbon nitride has a high ion tolerance (higher than any other material reported except PHI), which we tentatively attribute to its textural porosity, enabling ionic permeation by an internal flow of ions and liquid under illumination.[43, 44] However, at 1 M NaCl, the melon microswimmers stopped swimming, which may be attributed to the absence of optoionic properties and a smaller thermodynamic driving force for oxidation (**Figure 1d**).

In contrast to melon, PHI has not only textural, but also structural nanopores with a radius on the order of 3.84 Å.[45, 46, 48] The pores are large enough to host and allow the passage of hydrated $K^+$ and $Na^+$ ions (hydrodynamic radius of 1.25 and 1.84 Å, respectively), making the molecular backbone of PHI permeable to ions. We have demonstrated in earlier work that the light-induced hole extraction and photo-charging ability of PHI is intimately linked to the presence of ions in the pores of PHI and the surrounding electrolyte, which may lead to light-induced ion transport throughout the structural and textural nanopores of PHI, leading to the movement of ions towards the (photogenerated) electrons on PHI,[43, 46, 47, 49] as illustrated in **Figures 2g** and **2h**. We believe it is these intrinsic optoionic properties of PHI, which are absent in melon, paired with porosity, that lead to sustained motion in high ionic strength media on the timescale of the experiments. Both the presence of textural and structural pores invalidates the assumption of a hard sphere (the latter excluding liquid or ion flow through the volume of the particle), which is commonly used to describe phoresis. Hence the Helmholtz-Smoluchowski equation, $U=\mu_e E_0$, where $\mu_e$ is the electrophoretic mobility and $E_0$ is the magnitude of the electric field, cannot be used to sufficiently describe the motion. Therefore, new theoretical models need to be developed in the future to fully capture the reason for the high ion tolerance of this organic semiconductor.

Despite the presence of ions and no dedicated fuel, the speed of light-driven PHI microswimmers (15-23 µm/s) is comparable with, and even higher, than the speed of other light-driven swimmers in absence of ions and in presence of dedicated fuels (5-25 µm/s).[56, 57] The limitations on phoresis arising from the accumulation of ions around the particle are shown to be efficiently bypassed by suitable porosity and additionally, it is hypothesized that optoionic effects enabling ion transport increases the ionic tolerance and speed of the swimmers. In presence of buffers and biological fluidic media, the additional species partially negate the speed reduction, presumably by offering additional redox reaction pathways with respect to water and NaCl, hence increasing the efficiency of the photocatalytic propulsion process.

**Phototaxis of PHI microswimmers**

Light-induced propulsion itself lacks directional control. However, phototactic properties of the microswimmers and light illumination direction control enable their steering to a desired location, which is very beneficial for their biomedical and other practical applications. To test possible phototactic properties of the PHI particles, they were illuminated with 365 nm UV light at an 45º angle from one side, with an intensity of 115±15 mW/cm$^2$, in DMEM medium. This created a light gradient along the *x*-axis as shown in **Figure 2d**, which the particles followed. The mean speed of the microswimmers was 12.5 ± 0.4 µm/s in this case, which is significantly higher (78%) compared to the lateral motion measurement with the perpendicular illumination through the microscope objective at 400 mW/cm² (6.5 ± 1.2 µm/s), and hence as fast as with 10x stronger illumination from the bottom (1.2 W/cm²). This apparent increase in speed is attributed to a change in directional component of the measured speed. In the case of illumination from the side, there was a larger parallel component resulting in a higher lateral speed, whereas when illuminating from the bottom, the parallel component was slower, while the *z*-component (parallel to the illumination) was not analyzed, resulting in a lower effective speed. Besides, tumbling and rotational motion may cause a sidewise motion with the perpendicular illumination.[58] As can be seen from the trajectories in **Figure 2d**, the polar plot in **Figure 2e**, and Movie **S4**, the PHI particles followed the direction of illumination very precisely. When the direction of illumination was changed, the direction of swimming also changed, as shown in Movie **S4**. This behavior could be explained by the illumination from one side only, shadowing the other half of the particle, resulting in the creation of an artificial Janus-type structure that breaks the miscroswimmer symmetry. The negative zeta potential of PHI (-40 mV)[45] led to positive phototaxis in this situation. [22, 59, 60, 61]

When PHI microswimmers were illuminated under the optical microscope with light focused onto the image plane, the particles moved towards the middle of the light beam (**Movie S1** and **Figure 2f**), enabling collective assembly in one direction. Further effects and possible thermal contributions are discussed in **Supplementary Note 3** and **Figures S4-S6** showing control experiments with non-propelled, passive polystyrene (PS) particles. The phototactic properties of the PHI microswimmers, which originating predominantly from their light-induced photocatalytic propulsion (**Figures 2g** and **2h**), can be used to control the direction of their motion in various future biomedical applications, such as targeted drug delivery and cargo transport.

**Cytotoxicity tests**

A low cytotoxicity profile of the microswimmers is an essential requirement for their future biomedical applications. Therefore, we tested the cytotoxicity of the PHI microswimmers with a normal cell line (NIH 3T3 fibroblast cells) and two cancer cell lines (HT-29 colorectal cancer cells and SKBR3 breast cancer cells). Live/dead staining of the tested cell lines incubated with varying concentrations of PHI microswimmers (0-30 µg/mL) for 24 h showed no decrease in cell viability for all cell lines (**Figures 3a,b**). Other than particle cytotoxicity, we also tested the effect of illumination and catalytic activity of PHI microswimmers on cellular viability. Illumination at 415 nm (420 mW/cm$^2$) and 385 nm (1.1 W/cm$^2$) of high density of PHI microswimmers (30 µg/mL) on the fibroblast cells for varying durations (0-30 min) showed no negative effects on their viability, both immediately after testing and after 24 h incubation (**Figures 3c,d**). Coupled to the efficient propulsion of the PHI microswimmers in biological media, their operation in cellular environments is highly promising. When illuminated at 415 nm for 30 min, the cells retained viability up to 24 h. With 385 nm UV light, 24 h viability is retained up to 10 min of illumination. Since strong UV illumination for long durations is harmful for the cells,[62] they are not viable anymore if illumination lasts for 30 min. Studies with primary cells from mouse splenocytes further confirmed there was no detectable level of IL-12 (pro-inflammatory marker) in either the untreated samples in concentrations used above in the dark. As shown here, long-duration exposure to short wavelength light can cause damage to the cells; therefore, the swimmer exposure times need to be limited, which are 10 and 30 minutes for 415 and 385 nm wavelength light here, respectively. Such durations are typically long enough for the desired *in vitro* or *in vivo* biomedical functionalities.

**Drug loading and hypoxically, pH- and light-triggered drug release**

Biocompatible PHI microswimmers are capable of actively carrying and releasing drugs and other cargos attached to their porous body structure at a target location. The concept of their targeted drug delivery is illustrated in **Figure 4a** with an anti-cancer model drug, doxorubicin (DOX). The presence of textural nanopores in both PHI and melon microswimmers is beneficial for enhanced drug loading efficiency (Brunauer–Emmett–Teller (BET) surface area is ~13 and 26 m²/g and pore size distribution is from 5-20 nm and 5-40 nm for the PHI and melon microswimmers, respectively; see **Figure S7** and **Supplementary Note 4**), enabling the adsorption of drugs within the microparticle pores.[45, 63] We further anticipate that drug uptake is assisted by the amine surface groups of both carbon nitrides, enabling hydrogen bonding interactions with the drug, and by the negative zeta potential of melon and especially PHI, which intrinsically attracts the positively charged DOX molecules at pH 7.[63, 64, 45] To test this hypothesis, 200 µg of DOX was added to a suspension of 100 µg of PHI microswimmers dispersed in 1 mL DMEM, resulting in 185 µg DOX encapsulated with a DOX loading efficiency of 185% on PHI after 24 hours, which is far higher than the previously reported values of 20-70%.[65, 66, 67] **Figure 4b** shows the fluorescent image of DOX loaded on the PHI particles. For melon, a loading efficiency of 110% (110 µg) resulted under the same conditions. Astonishingly, no passive release was observed for PHI in the absence of illumination for more than 30 days. The stronger attachment and higher loading of DOX to the surface of PHI is likely due to enhanced interactions between the oxygen-rich DOX backbone and PHI, due to its more negative zeta potential in comparison to melon.[45, 46, 64] Moreover, the DOX molecules loaded on PHI particles had no negative effect on the propulsion speed of the PHI microswimmers in DMEM (18.5 ± 0.9 µm/s). This can be rationalized by the fact that DOX molecules are predominantly physisorbed in the particle volume, rather than on the outer particle surface, which does not significantly affect the outer surface photocatalytic reactions of PHI with the surrounding medium that give rise to a field gradient around the particle and hence, swimming propulsion.

Illuminating the PHI and melon particles triggers the release of DOX, thus enabling a fully controlled, on-demand release of the drug. We used a 415 nm blue light at an intensity of 170 mW/cm$^2$ to study the release of DOX and related products from the microswimmers in both ambient and O$_2$-free environments (**Figures 4c** and **S9a**). Band gap illumination of PHI microswimmers in oxygen-deficient suspensions triggers (oxidative) photocharging of the material, which is accompanied by a change of optoelectronic properties and a color change from yellow to blue.[48, 50] Hence, this charging effect is expected to influence the release too (**Figure**

**4a**). Interestingly, the charging effect was also observed for denser PHI suspensions in DMEM even in ambient conditions (i.e., in the presence of oxygen), which appears to originate from photocatalytic oxygen consumption near the PHI surface (**Figure S9b** and **Supplementary Note 5**). The cumulative release signal of DOX from PHI in DMEM is shown in **Figure 4d** for 3, 10 and 30 minute intervals of illumination in both ambient and $O_2$-free environments. Results of the DOX release with continuous illumination are shown in **Figure S10**. In ambient conditions, an optical equivalent of approximately 26 µg (14%) was released every 10 min, with 64 µg (35%) in 30 min cumulatively. When the amount of oxygen is decreased by purging the suspension with Ar 5 min prior and during the illumination, enabling photocharging of PHI, an increase (almost doubling) in stepwise and cumulative release was observed, resulting in 114.8 ± 5.2 µg (62%) of DOX equivalently being released after 30 min (**Figures 4c,d**). Tumor cells have oxygen-deficient, i.e. hypoxic, regions. Hence, a microswimmer which releases drugs faster in oxygen-deficient conditions is not only beneficial, but could even be used in a theranostic sense for hypoxically triggered drug delivery in tumor regions.

For melon, a light-triggered and rather constant DOX release was also observed after 3 and 10 min of illumination, albeit at lower overall amounts than for PHI, albeit at similar proportions (~17 mg of DOX equivalent, 15% of loading and hence significantly less than 67% release possible with PHI despite higher initial loading), and with no significant differences in oxygen-depleted conditions for these short time scales (**Figure S11**).

Upon longer illumination however, beyond 60 min illumination time intervals, the DOX amount released from PHI was reduced in ambient ($O_2$-rich) conditions, suggesting a light-triggered degradation process, similar to observations made on DOX-loaded $TiO_2$ microparticles.[68] Since degraded products of DOX show similar fluorescence and absorption bands, a further examination of the release products was only possible qualitatively by separation and mass spectroscopy, which was realized by high-performance liquid chromatography-mass spectroscopy (HPLC-MS) analysis of the supernatants after release. Besides pristine DOX, modified DOX byproducts were indeed observed after release (**Figures 4e** and **S12**). The dominant decomposition products have molecular masses of 413, 399 and 143 g/mol (see **Scheme S1** for visualization of the reaction; more information on the drug release behavior is discussed in **Supplementary Note 6**). This clearly gives the evidence that not only the release of DOX itself, but also of its oxidation products, which can act even more effectively on cells,[68] can be intentionally tuned by the illumination time while being responsive to the presence of oxygen.

Besides, the PHI microswimmers are sensitive to acidic pH levels. A low pH triggers protonation of the PHI backbone and an increase in zeta-potential, hence repelling the DOX.[46, 64] Within 60 min at pH 3.5, induced by adding HCl to PBS, the microswimmers released the loaded DOX (116.2 ± 3.3 µg, 65%) without any light illumination (control) as shown in **Figure 4f**. In comparison, the control (pH 6.7) showed negligible release of ~2%, within the margins of instrumental error.

Next, the ability of PHI microswimmers to deliver DOX or its active modifications to cancer cells under illumination was studied to provide a proof of concept. The bright field image of PHI microswimmers in a dense environment of cancer cells after 20 min of illumination (415 nm) is shown in **Figure 4g**. DOX or related products (red color) were released from the PHI microswimmers and taken up by the cancer cells after 24 h incubation, while some amounts also diffused through the medium. Subsequently, the incubated cells died. The fluorescent image of the PHI microswimmers shows that the microswimmers stayed in the vicinity of the cancer cells (**Figure 4h**). [69] This demonstrates that PHI microswimmers serve as smart light-driven cargo delivering agents in biological conditions. In fact, their stimuli-responsive behavior triggers a theranostic function, making use of an intrinsic sensing property (i.e., charge accumulation enabled in oxygen-poor environments) that triggers an *enhanced* release of the therapeutic agents.

**Conclusion**

We have developed highly efficient carbon-based PHI microswimmers that can be propelled phototactically with light in aqueous salt media with high molarity (up to 1 M NaCl) and realistic biological cell environments, such as dPBS, DMEM, DMEM/FBS, as well as diluted blood, without any additional fuel. Such particles exhibit positive phototaxis, which can be used to steer and locate the microswimmers into target locations,[18] and can be photocharged even in oxygen-rich conditions. Their high ion tolerance for phototaxis (>1 M) is attributed to a favorable interplay between PHI's textural and structural porosity, as well as possible optoionic effects enabling ion motion into and through the particle, in the presence of high photocatalytic activity.[48, 49] Next, the biocompatibility of the PHI swimmers was verified with three different cell lines without and with visible light illumination and primary cells. Besides, due to their textural porosity, PHI microswimmers are shown to have a very high capacity for drug uptake (~185% of their own mass using the example of the anti-cancer drug DOX), which stays bound stably to the particle over a month, until a fast release is triggered by a pH change or band-gap illumination. Intriguingly, PHI shows stimuli responsive drug release, as the release can be enhanced or modified by the intrinsic memristive photocharging ability in oxygen-deficient conditions, which are prevalent in hypoxic tumor regions, thus enabling potential future neuromorphic or theranostic applications.[49, 70, 71, 72, 73] In addition to the multi-stimuli responsive drug release capability of the PHI microswimmers, we demonstrate their swimming and ease of tracking in crowded heterogeneous biological fluids, such as diluted blood, by monitoring PHI's inherent autofluorescence. These capabilities make the PHI microswimmers promising candidates for future *in-vitro* and *in-vivo* biomedical applications in hypoxic tumor regions inside the human body and organ-on-a-chip devices. Light penetration into tissues is a challenge for all previous and our light-driven microswimmers for *in vivo* deep-tissue medical applications since short wavelength light cannot penetrate the tissue. To overcome this challenge, long wavelength near-infrared (NIR) light-driven microswimmers need to be developed in the future, which could penetrate several centimeters deep into the tissues. Due to PHI's organic nature, we envision that the microswimmers can be further optimized for tailored surface functionalities as well as catalytic, morphological and optical properties,[51,46] opening up new avenues for smart responsive micromachines and theranostics in the future.

**Materials and Methods**

**PHI synthesis.** PHI was synthesized according to a procedure described in the literature.[45] Shortly, melamine (5.0 g) is heated in a tube furnace in a quartz glass boat to 550°C for 12 h with a heating rate of 5°C/min under Argon (Ar) flow. After cooling to ambient temperature, a yellow powder (2.0-2.5 g) is obtained. 1.5 g of this product (melon) were thoroughly ground with KSCN (3.0 g), which was heated overnight to 140°C in vacuum to evaporate water. The mixture is heated in a tube furnace in an Alox boat to 400°C for 1 h and 500°C for 30 min with a heating rate of 30°C/min under Ar flow. The Alox boat with the $CN_x$ is sonicated two times for 15 min in 80 mL of water to disperse the yellow product. This suspension is washed six times with DI water by centrifugation (20,000 rpm). The insoluble product is dried in vacuum at 60 °C overnight.

SEM images of the microswimmers were captured by a Zeiss Merlin SEM. In order to capture the swimming of the microswimmers, a Zeiss Axio A1 inverted optical microscope was used. A Thorlabs M365L2 (Germany) UV lamp, Zeiss Colibri fluorescence source and Thorlabs M415 was used for illumination through the inverted microscope. The videos were recorded from the microscope using a LD Plan-NeoFluar 40x objective lens and Axiocam 503 CCD camera at 62 frames per second. The swimming behaviour of the microswimmers was systematically investigated by 2D mean-square-displacement (MSD) analysis on the captured videos for 15 s using a custom MATLAB and Python code.[1]

The absorptance spectra of these samples were measured with a double monochromator spectrophotometer (Edinburgh Instruments, FLS-980). The measurements were performed locating the sample in the centre of an integrating sphere attached to FLS-980 working in synchronous mode to discern any photoluminescence signal from the PHI particle. The sample suspension was measured in degassed, aqueous solution while being stirred to prevent sedimentation.

The intensity of the illumination in the microscope was measured at the place of the sample chamber with a calibrated Ocean Optics OCEAN-FX-XR1-ES spectrophotometer after attenuation by a neutral density filter and integration of the spectral irradiance. The results have been normalized to the filter attenuation and to the spot size of the light beam in the microscope, which was measured to be $2.0 \pm 0.5$ mm in diameter, resulting in a relative experimental error of 50% after the error propagation calculation. For side illumination with the 365 nm diode, the light intensity was directly measured by a Thorlabs S310C/PM100D power meter.

**Cytotoxicity tests.** Human colorectal cancer cells, human breast cancer cells, and murine fibroblasts (ATCC, Manassas, VA) were grown in DMEM supplemented with 10% v/v FBS and 1% v/v penicillin/streptomycin (Gibco, Grand Island, NY, USA) at 37 °C in a 5% CO2, 95% air humidified atmosphere. Cells were re-seeded after growing to confluence into µ-Slide 8 Well plates (Ibidi GmbH, Gräfelfing, Germany) at a cell density of $25 \times 10^3$ cells/well and incubated for 2 days. For cytotoxicity testing, all three cell lines were incubated with PHI microswimmers at varying concentrations (0-30 µg/mL). Then, viability of difference cell lines was investigated using a Live/Dead assay (ThermoFisher Scientific, Waltham, MA) incorporating calcein-AM (green) and ethidium homodimer-1 (red) dyes. Following 24 h of incubation with the PHI microswimmers, live/dead cell viability was calculated from fluorescence microscopy images. Furthermore, cytotoxicity of microswimmers during light actuation (385 nm and 415 mW/cm$^2$) was tested by live/dead staining of fibroblast cells following right after and 24 h after actuation of PHI microswimmers for varying durations (0-10 min).

**Cell culturing.** All cell culturing was performed under sterile conditions within a biosafety cabinet. The cell culture used for the isolated cells was composed of Dulbecco's Modified Eagle Medium (DMEM) from Gibco. This was supplemented with 10% Heat-inactivated Fetal Bovine Serum and 1% Penicillin and Streptomycin (Gibco). The storage condition for cells during incubation was an incubator set at 5% $CO_2$, ~90% humidity and 37°C.

**Cell isolation.** Mouse spleen was provided by the Facility for Animal Welfare, Veterinary Service and Laboratory Animal Science at Eberhard Karl's University Tubingen. After mice were sacrificed the spleen was removed and keep on ice in PBS without $Ca^{2+}/Mg^{2+}$ for transport to Max Planck Institute at Stuttgart. The spleen was forced through a cell strainer at 70 µm also containing DMEM media. After filtering, the supernatant was centrifuged at 800 x G for 5 min. The upper layer of supernatant was removed and the cells were washed with lysing buffer for red blood cells removal. This occurred for 8 minutes at room temperature. Then the cells were centrifuged and resuspended in fresh medium before being placed into 6-well plates at 1 million cells per well. The chosen concentrations of the tested material were added to the wells for 24 hours before the medium was removed and frozen before Elisa analysis.

**Elisa protocol (IL-12).** A high affinity binding plate (Greiner) was coated with 100 µl of diluted, purified anti-cytokine capture antibody. The plate was sealed and incubated at 4°C overnight. The next day the capture anti-body was removed by decanting and a blocking solution was added for 1 hour at room temperature. A series of washes was performed between each step. The samples and standards were added and incubated for 2 hours at room temperature. After rinsing the working detector was added and incubated for 1 hour. The final rinse was performed and then the substrate solution was added and the plate was read after adding stop solution at 30 minutes.

**Details of swimming in biofluids.** DMEM, dPBS and DMEM+FBS (Sigma) were used as purchased. The PHI swimmers were dispersed in the medium and illuminated inside a microfluidic chamber to measure their swimming.

**Details of DOX loading.** The efficiency of loading was measured by centrifuging the DOX loaded microswimmers and comparing the optical density (OD) of the supernatant with the precalibrated OD of 200 µg/ml DOX at 480 nm. 100 µg/ml of PHI was dispersed with 200 mg/ml of DOX and this solution was stirred in dark for 24 hours to allow the DOX to be adsorbed. After 24 hours the suspension was centrifuged and the supernatant was used for measuring the DOX loading. The DOX loaded PHI was washed three times with water and stored at 4 °C for further delivery experiments. The DOX loaded microswimmers were illuminated with 415 nm light at an intensity of 170 mW/cm$^2$ from the bottom in a cylindrical quartz glass while stirring (**Figure S9a**). To remove dissolved oxygen, the suspension was bubbled with argon through a needle for 5 min prior to the illumination and during the respective illumination time. For the pH release the pH of the resulting HCl diluted PBS solution was checked using a pH meter to confirm the stability of the pH during the release experiments.

**Sorption measurements.** Sorption measurements were performed on a Quantachrome Instruments Autosorb iQ 3 with a coupled Cryosync for cooling and Argon as analysis gas at 87K. The pore size distribution was determined from argon adsorption isotherms using the NLDFT (cylindrical nanopores, adsorption branch) kernel in carbon for argon at 87 K implemented in the ASiQwin software v 3.01. Samples were activated in high vacuum at 120 °C for 12 h before measurement.

**High-performance liquid chromatography-mass spectroscopy (HPLC-MS) analysis.** High pressure liquid chromatography with mass spectrometry (HPLC-MS) was performed on an Agilent 1290 Infinity II LC system connected to an Agilent InfinityLab LC/MSD XT single quadrupole mass spectrometer with a multimode ESI-APCI ionization source. Analysis of the combined signals was performed using MestReNova (Version 14) software package with MS analysis tools. Chromatographic separation was achieved on an Agilent EclipsePlus C18 column (2.1 x 50 mm, 1.8 µm) at 40°C with mixtures of acetonitrile (MeCN), water and formic acid (FA), according to the solvent composition timetable (Tables S5 and S6) and a total solvent flow of 0.7 mL/min. MS data was obtained using MM-APCI ionization (positive) and in selective ion monitoring mode for signals m/z = 544.2 (DOX), 399.1, 413.1 and 148.1 (degradation products).

*Sample preparation*: After centrifugation of the particles, an aliquot of the supernatant (100 µL) was diluted with 400 µL MeCN:water (80:20 v/v). The diluted samples were filtered through a syringe filter (0.2µm, PTFE) and injected (5 µL) into the HPLC-MS.

*Calibration*: In order to determine the concentration of doxorubicin quantitatively, a multilevel calibration of the mass signal (area, SIM) was performed with a concentration series (Table S5 and Figure S13). An appropriate volume of stock solution of doxorubicin (c = 1 mg/mL in water) was diluted with D-MEM to prepare the samples $C_1$-$C_4$. Then, an aliquot (15 µL) of the calibration sample was diluted with 985 µL MeCN:water (80:20 v/v). The diluted samples were filtered through a syringe filter (0.2µm, PTFE) and injected (1 µL) into the HPLC-MS.

*Quantification of DOX*: The concentration of DOX in an unknown sample was calculated from the peak area of the mass trace measured in selective ion monitoring (SIM) mode (m/z = 544.18) as follows (see **Figure S13-S15** for calibration curve and chromatograms):

$$c(DOX) = \frac{(Area - 757.48508)}{42517.78309} \frac{\mu g}{mL}$$

Note that this equation already includes the sample dilution factor (0.2) and injection volume (5 µL).

**Acknowledgements**

We thank Dr. Yan Yu for her help with cell culture and imaging, Viola Duppel for SEM imaging, and Sebastian Emmerling for BET measurements. Financial support by the Max Planck Society, the European Research Council (ERC) Advanced Grant SoMMoR project with grant no: 834531, the ERC Starting Grant COFLeaf project with grant no: 639233, the Deutsche Forschungsgemeinschaft (DFG) via the cluster of excellence "e-conversion" (project number EXC2089/1–390776260), and by the Center for NanoScience (CENS) is gratefully acknowledged.

**Conflict of interest**

The authors declare no conflict of interest.

**Author contributions**

V.S. and F.P. as well as M.S., J.K., B.L. conceived and planned the research. V.S., F.P., Y.A., J.K., L.G. and V.K. performed the experiments or assisted in their realization. All authors contributed to the analysis and discussion of the data. M.S., F.P. and B.L. supervised the research. V.S. and F.P. wrote the manuscript with assistance from all other authors.

**Data availability**

The raw data is available from the corresponding authors upon reasonable request.

# References

1. M.Sitti. Mobile microrobotics. *MIT Press, Cambridge, MA* 2017.

2. Sanchez S, Soler L, Katuri J. Chemically powered micro- and nanomotors. *Angew Chem Int Ed Engl* 2015, **54**(5)**:** 1414-1444.

3. Erkoc P, Yasa IC, Ceylan H, Yasa O, Alapan Y, Sitti M. Mobile Microrobots for Active Therapeutic Delivery. *Advanced Therapeutics* 2019, **2**(1)**:** 1800064.

4. Eskandarloo H, Kierulf A, Abbaspourrad A. Light-harvesting synthetic nano- and micromotors: a review. *Nanoscale* 2017, **9**(34)**:** 12218-12230.

5. Alapan Y, Yasa O, Yigit B, Yasa IC, Erkoc P, Sitti M. Microrobotics and Microorganisms: Biohybrid Autonomous Cellular Robots. 2019, **2**(1)**:** 205-230.

6. Luo M, Feng Y, Wang T, Guan J. Micro-/Nanorobots at Work in Active Drug Delivery. 2018, **28**(25)**:** 1706100.

7. Ma X, Sánchez S. Self-propelling micro-nanorobots: challenges and future perspectives in nanomedicine. *Nanomedicine* 2017, **12**(12)**:** 1363-1367.

8. Li J, Esteban-Fernández de Ávila B, Gao W, Zhang L, Wang J. Micro/nanorobots for biomedicine: Delivery, surgery, sensing, and detoxification. *Science Robotics* 2017, **2**(4)**:** eaam6431.

9. de Ávila BE-F, Angsantikul P, Li J, Angel Lopez-Ramirez M, Ramírez-Herrera DE, Thamphiwatana S*, et al.* Micromotor-enabled active drug delivery for in vivo treatment of stomach infection. *Nature Communications* 2017, **8**(1)**:** 272.

10. Wu Z, Li L, Yang Y, Hu P, Li Y, Yang S-Y*, et al.* A microrobotic system guided by photoacoustic computed tomography for targeted navigation in intestines in vivo. *Science Robotics* 2019, **4**(32)**:** eaax0613.

11. Wang J, Gao W. Nano/Microscale Motors: Biomedical Opportunities and Challenges. *ACS Nano* 2012, **6**(7)**:** 5745-5751.

12. Gao W, Dong R, Thamphiwatana S, Li J, Gao W, Zhang L*, et al.* Artificial Micromotors in the Mouse's Stomach: A Step toward in Vivo Use of Synthetic Motors. *ACS Nano* 2015, **9**(1)**:** 117-123.

13. Ma X, Hortelão AC, Patiño T, Sánchez S. Enzyme Catalysis To Power Micro/Nanomachines. *ACS Nano* 2016, **10**(10)**:** 9111-9122.

14. Llopis-Lorente A, García-Fernández A, Murillo-Cremaes N, Hortelão AC, Patiño T, Villalonga R*, et al.* Enzyme-Powered Gated Mesoporous Silica Nanomotors for On-Command Intracellular Payload Delivery. *ACS Nano* 2019, **13**(10)**:** 12171-12183.

15. Hortelão AC, Patiño T, Perez-Jiménez A, Blanco À, Sánchez S. Enzyme-Powered Nanobots Enhance Anticancer Drug Delivery. *Advanced Functional Materials* 2018, **28**(25)**:** 1705086.

16. Šípová-Jungová H, Andrén D, Jones S, Käll M. Nanoscale Inorganic Motors Driven by Light: Principles, Realizations, and Opportunities. *Chemical Reviews* 2020, **120**(1)**:** 269-287.

17. Xu L, Mou F, Gong H, Luo M, Guan J. Light-driven micro/nanomotors: from fundamentals to applications. *Chemical Society Reviews* 2017, **46**(22)**:** 6905-6926.

18. Wang J, Xiong Z, Zheng J, Zhan X, Tang J. Light-Driven Micro/Nanomotor for Promising Biomedical Tools: Principle, Challenge, and Prospect. *Acc Chem Res* 2018, **51**(9)**:** 1957-1965.

19. Zhang D, Sun Y, Li M, Zhang H, Song B, Dong B. A phototactic liquid micromotor. *Journal of Materials Chemistry C* 2018, **6**(45)**:** 12234-12239.

20. Kong L, Mayorga-Martinez CC, Guan J, Pumera M. Photocatalytic Micromotors Activated by UV to Visible Light for Environmental Remediation, Micropumps, Reversible Assembly, Transportation, and Biomimicry. *Small* 2019**:** e1903179.

21. Uspal WE. Theory of light-activated catalytic Janus particles. 2019, **150**(11)**:** 114903.

22. Dai B, Wang J, Xiong Z, Zhan X, Dai W, Li CC*, et al.* Programmable artificial phototactic microswimmer. *Nat Nanotechnol* 2016, **11**(12)**:** 1087-1092.

23. You M, Chen C, Xu L, Mou F, Guan J. Intelligent Micro/nanomotors with Taxis. *Acc Chem Res* 2018, **51**(12)**:** 3006-3014.

24. Kuron M, Kreissl P, Holm C. Toward Understanding of Self-Electrophoretic Propulsion under Realistic Conditions: From Bulk Reactions to Confinement Effects. *Accounts of Chemical Research* 2018, **51**(12)**:** 2998-3005.

25. Moran JL, Posner JD. Role of solution conductivity in reaction induced charge auto-electrophoresis. 2014, **26**(4)**:** 042001.

26. Brown A, Poon W. Ionic effects in self-propelled Pt-coated Janus swimmers. *Soft Matter* 2014, **10**(22)**:** 4016-4027.

27. Sridhar V, Park B-W, Guo S, van Aken PA, Sitti M. Multiwavelength-Steerable Visible-Light-Driven Magnetic CoO–TiO2 Microswimmers. *ACS Applied Materials & Interfaces* 2020, **12**(21)**:** 24149-24155.

28. Sridhar V, Park B-W, Sitti M. Light-Driven Janus Hollow Mesoporous TiO2–Au Microswimmers. 2018, **28**(25)**:** 1704902.

29. Wang X, Sridhar V, Guo S, Talebi N, Miguel-López A, Hahn K*, et al.* Fuel-Free Nanocap-Like Motors Actuated Under Visible Light. 2018, **28**(25)**:** 1705862.

30. Wang X, Baraban L, Nguyen A, Ge J, Misko VR, Tempere J*, et al.* High-Motility Visible Light-Driven Ag/AgCl Janus Micromotors. *Small* 2018, **14**(48)**:** e1803613.

31. Wang J, Xiong Z, Zhan X, Dai B, Zheng J, Liu J*, et al.* A Silicon Nanowire as a Spectrally Tunable Light-Driven Nanomotor. 2017, **29**(30)**:** 1701451.

32. Villa K, Novotny F, Zelenka J, Browne MP, Ruml T, Pumera M. Visible-Light-Driven Single-Component BiVO4 Micromotors with the Autonomous Ability for Capturing Microorganisms. *ACS Nano* 2019, **13**(7)**:** 8135-8145.

33. Gibbs JG. Shape- and Material-Dependent Self-Propulsion of Photocatalytic Active Colloids, Interfacial Effects, and Dynamic Interparticle Interactions. *Langmuir* 2019.

34. Zhan X, Wang J, Xiong Z, Zhang X, Zhou Y, Zheng J*, et al.* Enhanced ion tolerance of electrokinetic locomotion in polyelectrolyte-coated microswimmer. *Nat Commun* 2019, **10**(1)**:** 3921.

35. Moran JL, Posner JD. Role of solution conductivity in reaction induced charge auto-electrophoresis. *Physics of Fluids* 2014, **26**(4)**:** 042001.

36. Wei M, Zhou C, Tang J, Wang W. Catalytic Micromotors Moving Near Polyelectrolyte-Modified Substrates: The Roles of Surface Charges, Morphology, and Released Ions. *ACS Applied Materials & Interfaces* 2018, **10**(3)**:** 2249-2252.

37. Palacci J, Sacanna S, Kim SH, Yi GR, Pine DJ, Chaikin PM. Light-activated self-propelled colloids. *Philos Trans A Math Phys Eng Sci* 2014, **372**(2029).

38. Wang Q, Dong R, Wang C, Xu S, Chen D, Liang Y, *et al.* Glucose-Fueled Micromotors with Highly Efficient Visible-Light Photocatalytic Propulsion. *ACS Applied Materials & Interfaces* 2019, **11**(6)**:** 6201-6207.

39. Aziz A, Pane S, Iacovacci V, Koukourakis N, Czarske J, Menciassi A*, et al.* Medical Imaging of Microrobots: Toward In Vivo Applications. *ACS Nano* 2020, **14**(9)**:** 10865-10893.

40. Lau VW-h, Mesch MB, Duppel V, Blum V, Senker J, Lotsch BV. Low-Molecular-Weight Carbon Nitrides for Solar Hydrogen Evolution. *Journal of the American Chemical Society* 2015, **137**(3)**:** 1064-1072.

41. Wang X, Maeda K, Thomas A, Takanabe K, Xin G, Carlsson JM*, et al.* A metal-free polymeric photocatalyst for hydrogen production from water under visible light. *Nature Materials* 2009, **8**(1)**:** 76-80.

42. Ong W-J, Tan L-L, Ng YH, Yong S-T, Chai S-P. Graphitic Carbon Nitride (g-C3N4)-Based Photocatalysts for Artificial Photosynthesis and Environmental Remediation: Are We a Step Closer To Achieving Sustainability? *Chemical Reviews* 2016, **116**(12)**:** 7159-7329.

43. Xiao K, Chen L, Chen R, Heil T, Lemus SDC, Fan F*, et al.* Artificial light-driven ion pump for photoelectric energy conversion. *Nature Communications* 2019, **10**(1)**:** 74.

44. Xiao K, Giusto P, Wen L, Jiang L, Antonietti M. Nanofluidic Ion Transport and Energy Conversion through Ultrathin Free-Standing Polymeric Carbon Nitride Membranes. *Angewandte Chemie International Edition* 2018, **57**(32)**:** 10123-10126.

45. Lau VW-h, Moudrakovski I, Botari T, Weinberger S, Mesch MB, Duppel V*, et al.* Rational design of carbon nitride photocatalysts by identification of cyanamide defects as catalytically relevant sites. *Nature Communications* 2016, **7**(1)**:** 12165.

46. Schlomberg H, Kröger J, Savasci G, Terban MW, Bette S, Moudrakovski I*, et al.* Structural Insights into Poly(Heptazine Imides): A Light-Storing Carbon Nitride Material for Dark Photocatalysis. *Chemistry of Materials* 2019, **31**(18)**:** 7478-7486.

47. Savateev A, Pronkin S, Willinger MG, Antonietti M, Dontsova D. Towards Organic Zeolites and Inclusion Catalysts: Heptazine Imide Salts Can Exchange Metal Cations in the Solid State. *Chemistry – An Asian Journal* 2017, **12**(13)**:** 1517-1522.

48. Podjaski F, Kroger J, Lotsch BV. Toward an Aqueous Solar Battery: Direct Electrochemical Storage of Solar Energy in Carbon Nitrides. *Adv Mater* 2018, **30**(9)**:** 1705477.

49. Podjaski F, Lotsch BV. Optoelectronics Meets Optoionics: Light Storing Carbon Nitrides and Beyond. *Advanced Energy Materials* 2021, **11**(4)**:** 2003049.

50. Lau VW-h, Klose D, Kasap H, Podjaski F, Pignié M-C, Reisner E*, et al.* Dark Photocatalysis: Storage of Solar Energy in Carbon Nitride for Time-Delayed Hydrogen Generation. *Angew Chem Int Ed Engl* 2017, **56**(2)**:** 510-514.

51. Kröger J, Jiménez-Solano A, Savasci G, Rovó P, Moudrakovski I, Küster K*, et al.* Interfacial Engineering for Improved Photocatalysis in a Charge Storing 2D Carbon Nitride: Melamine Functionalized Poly(heptazine imide). *Advanced Energy Materials* 2020, **n/a**(n/a)**:** 2003016.

52. Sridhar V, Podjaski F, Kröger J, Jiménez-Solano A, Park B-W, Lotsch BV*, et al.* Carbon nitride-based light-driven microswimmers with intrinsic photocharging ability. *Proceedings of the National Academy of Sciences* 2020**:** 202007362.

53. Pacheco M, Jurado-Sanchez B, Escarpa A. Visible-Light-Driven Janus Microvehicles in Biological Media. *Angew Chem Int Ed Engl* 2019, **58**(50)**:** 18017-18024.

54. Honn KV, Singley JA, Chavin W. Fetal Bovine Serum: A Multivariate Standard. *Proceedings of the Society for Experimental Biology and Medicine* 1975, **149**(2)**:** 344-347.

55. Singh M, Coulter NA. Rheology of blood: Effect of dilution with various dextrans. *Microvascular Research* 1973, **5**(2)**:** 123-130.

56. Qin H, Wu X, Xue X, Liu H. Light actuated swarming and breathing-like motion of graphene oxide colloidal particles. *Communications Chemistry* 2018, **1**(1)**:** 72.

57. Dai B, Wang J, Xiong Z, Zhan X, Dai W, Li C-C*, et al.* Programmable artificial phototactic microswimmer. *Nature Nanotechnology* 2016, **11**(12)**:** 1087-1092.

58. Uspal WE. Theory of light-activated catalytic Janus particles. *The Journal of Chemical Physics* 2019, **150**(11)**:** 114903.

59. Wang J, Wu H, Liu X, Liang Q, Bi Z, Wang Z*, et al.* Carbon-Dot-Induced Acceleration of Light-Driven Micromotors with Inherent Fluorescence. *Advanced Intelligent Systems* 2020, **2**(3)**:** 1900159.

60. Chen C, Mou F, Xu L, Wang S, Guan J, Feng Z*, et al.* Light-Steered Isotropic Semiconductor Micromotors. *Advanced Materials* 2017, **29**(3)**:** 1603374.

61. Sun Y, Jiang J, Zhang G, Yuan N, Zhang H, Song B, *et al.* Visible Light-Driven Micromotor with Incident-Angle-Controlled Motion and Dynamic Collective Behavior. *Langmuir* 2021, **37**(1)**:** 180-187.

62. Bryant SJ, Nuttelman CR, Anseth KS. Cytocompatibility of UV and visible light photoinitiating systems on cultured NIH/3T3 fibroblasts in vitro. *Journal of Biomaterials Science, Polymer Edition* 2000, **11**(5)**:** 439-457.

63. Gao Y, Chen Y, Ji X, He X, Yin Q, Zhang Z, *et al.* Controlled Intracellular Release of Doxorubicin in Multidrug-Resistant Cancer Cells by Tuning the Shell-Pore Sizes of Mesoporous Silica Nanoparticles. *ACS Nano* 2011, **5**(12)**:** 9788-9798.

64. Lau VW-h, Yu VW-z, Ehrat F, Botari T, Moudrakovski I, Simon T, *et al.* Urea-Modified Carbon Nitrides: Enhancing Photocatalytic Hydrogen Evolution by Rational Defect Engineering. *Advanced Energy Materials* 2017, **7**(12)**:** 1602251.

65. Schmidt CK, Medina-Sánchez M, Edmondson RJ, Schmidt OG. Engineering microrobots for targeted cancer therapies from a medical perspective. *Nature Communications* 2020, **11**(1)**:** 5618.

66. Wang B, Kostarelos K, Nelson BJ, Zhang L. Trends in Micro-/Nanorobotics: Materials Development, Actuation, Localization, and System Integration for Biomedical Applications. *Advanced Materials* 2020, **n/a**(n/a)**:** 2002047.

67. Dong J, Zhao Y, Wang K, Chen H, Liu L, Sun B, *et al.* Fabrication of Graphitic Carbon Nitride Quantum Dots and Their Application for Simultaneous Fluorescence Imaging and pH-Responsive Drug Release. *ChemistrySelect* 2018, **3**(44)**:** 12696-12703.

68. Calza P, Medana C, Sarro M, Rosato V, Aigotti R, Baiocchi C, *et al.* Photocatalytic degradation of selected anticancer drugs and identification of their transformation products in water by liquid chromatography–high resolution mass spectrometry. *Journal of Chromatography A* 2014, **1362:** 135-144.

69. Zhang W, Dang G, Dong J, Li Y, Jiao P, Yang M, *et al.* A multifunctional nanoplatform based on graphitic carbon nitride quantum dots for imaging-guided and tumor-targeted chemo-photodynamic combination therapy. *Colloids and Surfaces B: Biointerfaces* 2021, **199:** 111549.

70. Jo SH, Chang T, Ebong I, Bhadviya BB, Mazumder P, Lu W. Nanoscale Memristor Device as Synapse in Neuromorphic Systems. *Nano Letters* 2010, **10**(4)**:** 1297-1301.

71. Rivnay J, Inal S, Salleo A, Owens RM, Berggren M, Malliaras GG. Organic electrochemical transistors. *Nature Reviews Materials* 2018, **3**(2)**:** 17086.

72. Xiao K, Wan C, Jiang L, Chen X, Antonietti M. Bioinspired Ionic Sensory Systems: The Successor of Electronics. *Advanced Materials* 2020, **32**(31)**:** 2000218.

73. Pérez-Tomás A. Functional Oxides for Photoneuromorphic Engineering: Toward a Solar Brain. *Advanced Materials Interfaces* 2019, **6**(15)**:** 1900471.

# Figures

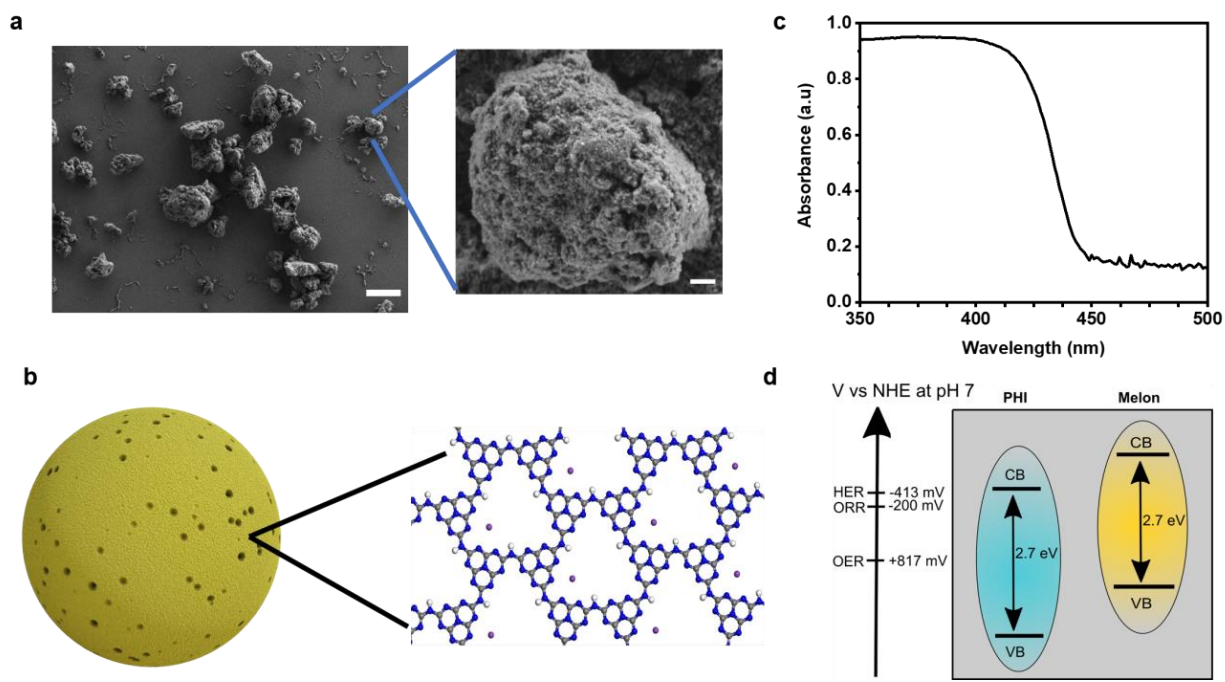

**Figure 1. Structure, morphology and optical properties of poly(heptazine imide) (PHI)-based organic microswimmer particles**. a) SEM image of representative PHI microparticles (grey) with a size distribution of 1-5 µm (scale bar: 5 µm) and close-up of one particle (scale bar: 400 nm), showing the porous morphology. b) Schematics of the PHI microswimmer and structure of the PHI macromolecules consisting of heptazine moieties comprising carbon (blue), nitrogen (grey) and hydrogen (white). Solvated potassium ions reside in structural nanopores (purple). c) Absorbance spectrum of PHI microswimmers, showing the onset of band gap absorption at 450 nm determined based on a Tauc plot. d) Band positions of PHI and melon with a bandgap of 2.7 eV along with hydrogen evolution reaction (HER), oxygen reduction reaction (ORR) and oxygen evolution reaction (OER) potentials at pH of 7.

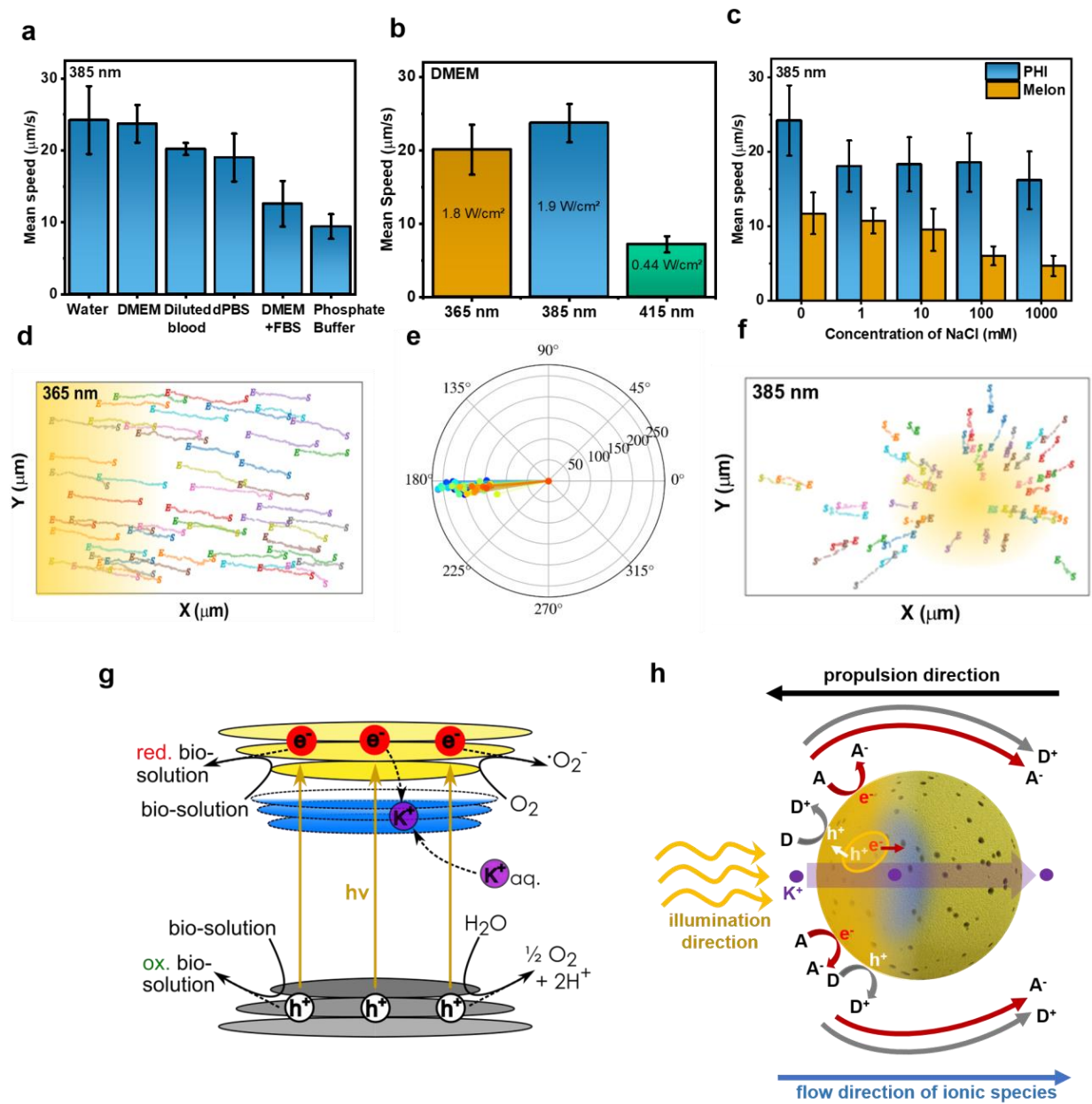

**Figure 2. PHI microswimmer propulsion in different ionic fluids.** a) Mean speed (N=50) of PHI microswimmers in various biological media swimming under 385 nm light illumination. b) Mean speed (N=50) of PHI microswimmers in DMEM medium under different illumination wavelengths. c) Mean speed (N=50) of the PHI and melon carbon nitride swimmers in different concentrations of NaCl under 385 nm illumination. d) Sample two-dimensional trajectory of the PHI microswimmers in DMEM medium swimming under 365 nm illumination from the side with S and E indicating the start and end of trajectories, showing a positive phototactic behavior. e) Polar plot of the directed propulsion of the PHI microswimmers in the same conditions as d). f) Phototactic behavior of PHI microswimmers in DMEM medium swimming under 385 nm illumination from the bottom to a point in the image. S and E indicate the start and end of the trajectories. g) Schematics of the band structure of PHI along with the reduction and oxidation reactions responsible for the photocatalytic propulsion of PHI microswimmers and the light-induced intercalation of alkali metal ions ($K^+$ or $Na^+$, e.g., purple) into the particle, enabled by optoionic effects assisting photocharging in PHI (blue). h) Proposed mechanism for the phototactic propulsion of porous carbon nitride particles (nanopores not to the real scale and density; particle shape not perfectly spherical in reality) caused by asymmetric illumination and photocatalysis, resulting in ion flow around and through the particle. The movement of cations (purple) into and through the particle's pores counteracts the Debye layer collapse and enables ionic tolerance. Pseudocapacitive photocharging effects (blue) present in PHI increase ionic tolearance with respect to melon.

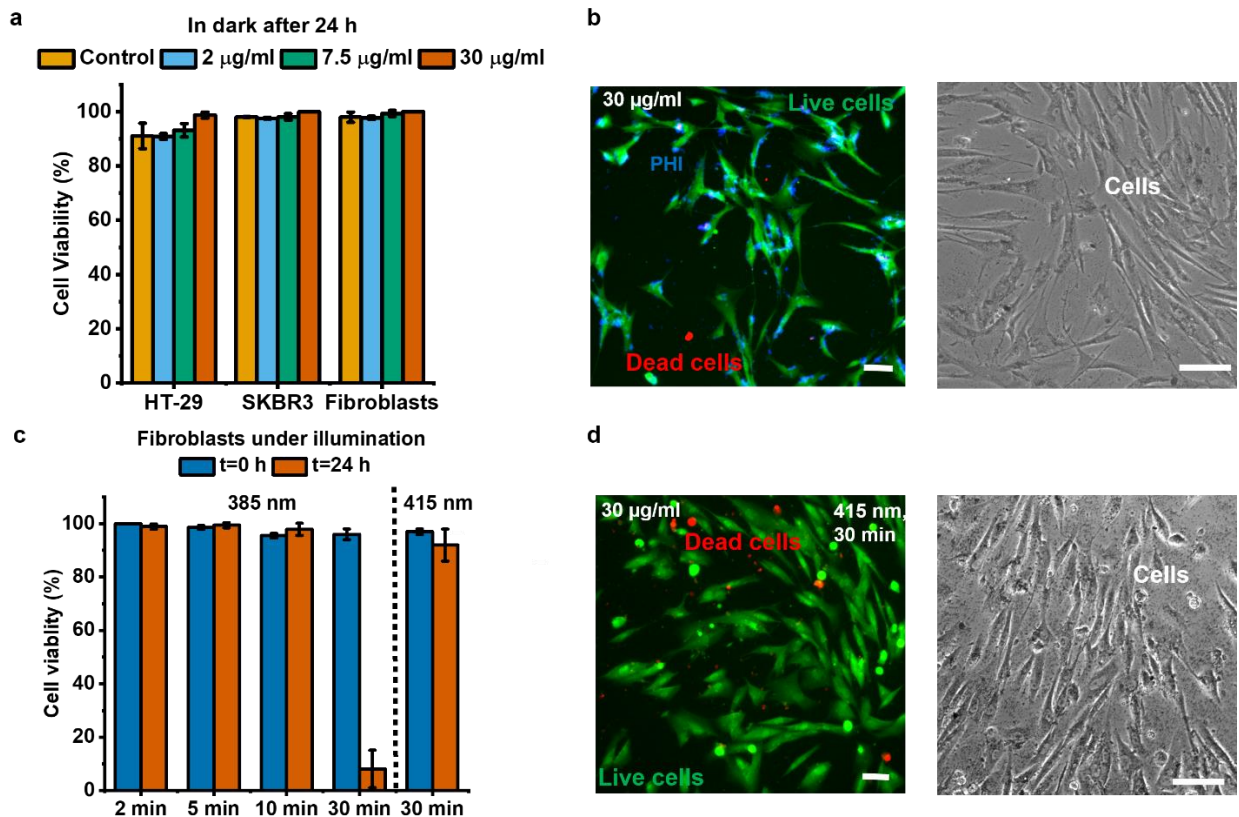

**Figure 3. Cytotoxicity of the PHI microswimmers.** a) Cell viability of fibroblasts and cancer cells incubated with the PHI microswimmers at varying concentrations after 24 hours. Data represents means ± standard deviation of ~300 cells. b) Live/dead staining of healthy BJ human fibroblast cells following 24 h incubation with the PHI microswimmers. Green and red indicates the live cells and dead cells, respectively. Along with the bright field images in the same conditions indicating the cells and PHI (black dots). Scale bar: 100 µm. c) Cell viability of BJ fibroblast cells in presence of PHI microswimmers (30 µg/ml) right after and 24 h after illumination for varying durations at 385 nm and 415 nm. Data represents means ± standard deviation. d) Live/dead staining of BJ fibroblast cells following swimming of PHI swimmers, not in the picture (30 µg/mL) via light (415 nm) after 30 mins. Along with the bright field images in the same conditions indicating the cells and PHI (black dots). Scale bar: 100 µm.

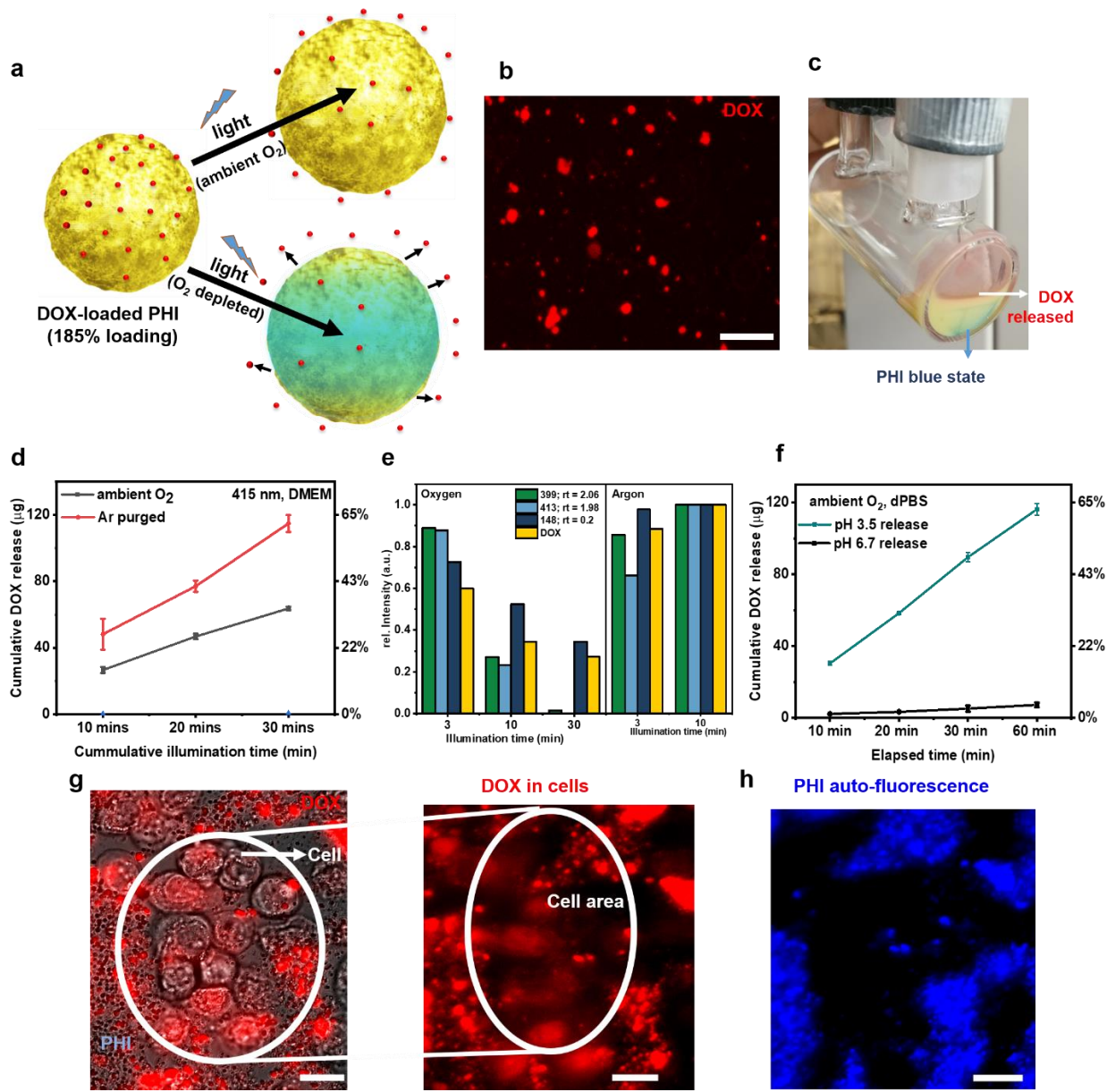

**Figure 4. Drug loading and hypoxically, light- and pH-triggered drug release with the PHI microswimmers.** a) Schematic drug loading (DOX) and triggered release from the PHI microswimmers in ambient and hypoxic conditions. b) Map of the DOX-loaded PHI microswimmers (185 µg/ml of DOX in 100 µg/ml PHI) showing DOX fluorescence emission at 595 nm. c) Photo of the DOX-loaded PHI microswimmers immediately after 30 min of illumination in oxygen-deficient conditions showing the charged blue state of PHI and the released DOX and byproducts in red color. d) Light-triggered cumulative release signal of DOX and byproducts in DMEM medium at different time points under 415 nm illumination, with the supernatant being removed at each interval in ambient (black) and Ar-purged (red). The non-cumulative release data is shown in **Figure S10**. e) HPLC analysis of the main products found after photocatalytic DOX release from the PHI microswimmer in ambient and Ar-purged conditions at different illumination times, normalized to the highest signal of the DOX that was observed from the HPLC (see **Supplementary Note 6** for details). f) pH-triggered release of DOX in PBS medium at pH 3.5 (green) with 54% of DOX being released after 60 min versus control (pH 6.7, black) showing negligible (~2%) release. g) Optical microscopy (bright field image) and fluorescence overlaid images of SKBR cancer cells (indicated within the circle) with pre-loaded PHI microswimmers under 415 nm illumination for 20 min showing released DOX uptake by the cells under emission at 595 nm fluorescence in red. h) PHI autofluorescence image at 470 nm showing the PHI microswimmers adhering to the cancer cells after some amount of DOX release. The two fluorescent images are taken from the same frame.